# Ship-induced solitons as a manifestation of critical phenomena

Stanyslav Zakharov[*] and Alexey Kryukov[*]

*A ship, moving with small acceleration in a reservoir of uniform depth, can be subjected to a sudden hydrodynamical impact similar to collision with an underwater rock, and on water surface unusual solitary wave will start running. The factors responsible for formation of solitons induced by a moving ship are analyzed. Emphasis is given to a phenomenon observed by John Scott Russell more 170 years ago when a sudden stop of a boat preceded the occurrence of exotic water dome. In dramatic changes of polemic about the stability and mathematical description of a solitary wave, the question why "Russell's wave" occurred has not been raised, though attempts its recreation invariably suffered failure. In our report the conditions disclosing the principle of the famous event as a critical phenomenon are described. In a reservoir of uniform depth a ship can confront by a dynamic barrier within narrow limits of ship's speed and acceleration. In a wider interval of parameters a ship generates a satellite wave, which can be transformed in a different-locking soliton. These phenomena can be classified into an extensive category of dynamic barrier effects including the transition of aircrafts through the sound barrier.*

A moving ship generates a sequence of surface gravity waves, but in special cases a single elevation with a stable profile can arise. It is the solitary wave, for the first time described by Russell in the form of almost a poem in prose (*1*). Similar, in a mathematical sense, the objects are now found everywhere from the microcosm up to the macrocosm (*2-6*), but water remains the most convenient and accessible testing ground for their studying. Earlier the water solitons were produced in a channel via a part of its bottom that was moved up impulsively (*7,8*) and the results of the experiments appeared to be in good agreement with theory (*9*). Initial conditions preceding the birth of the Russell's soliton (RS) were markedly different: a boat moved smoothly, and there were no visible reasons for its stop. Having lost the kinetic energy the boat emitted it in the form of a soliton. It is possible, this occurred once a dynamical parameter of the boat-canal system had attained a threshold value. The threshold conditions are usually simple, and this stimulated us to their search.

---

[*] Lebedev Physical Institute, Russian Academy of Sciences, Moscow.

We started with the reconstruction of some missing parameters assuming that the Union Canal near Edinburgh (*1*) had a horizontal bottom. Since the RS propagation obeys the Korteveg – de Vries's (KdV) equation (*10*) (the term is the Russell-KdV soliton) we used its solution to find the canal depth *H*. Speed of moving of a lonely wave on a surface of a reservoir The velocity of moving a solitary wave on a reservoir surface $v = \sqrt{gH}(1+\frac{h}{2H})$ with *h* as its height (*g* is the gravity acceleration). Substituting *v* and *h* values reported by Russell we determine the most probable limits for the *H* between 1.6 and 1.9 m (Fig. 1a).

We proposed that the phenomenon has been initiated by a ship wave, i.e. boat-generated steady-state gravity wave. Therefore, it is necessary to find a way to estimate the value of its wavelength $\lambda$. For this purpose we have applied a theory of the ship waves (*11*). The ship wave is formed of primary perturbations, and their spectrum depends on the speed of the ship. In the case under consideration the boat's speed is easily estimated. Assuming that the Russell's boat had the displacement about 50 ton and the length *L* of order 10 m its 'two-horsepower' motor hardly could outrun a walker. Hence, *V* is about 6 km per hour.

The ship wave is formed from a superposition of wave perturbations generated by a ship. In general, the perturbation spectrum $f(k)$ where $k = \frac{2\pi}{\lambda}$, may be wide. In our case, a condition $L > 2\pi \frac{V^2}{g}$ is valid and then, according (*11*), $f(k)$ is reduced to a narrow wave packet near the long-wavelength limit $k_l = \frac{g}{V^2}$. Therefore, the wavelength of the ship wave is uniquely determined as $\lambda_R \approx 2\pi \frac{V^2}{g}$, and its phase velocity *u* is equal to the boat's speed both by module and direction: $u = V\cos\vartheta$, $\vartheta = 0$. Substituting of numerical values gives $\lambda_R$ from 1.6 up to 2.0 m (Fig. 1b). A notable coincidence of λ and *H* supports the assumption on a threshold nature of the phenomenon.

As known, the surface gravity waves of various wavelengths are propagating with different velocities. Using the dispersion law $u^2 = \left(\frac{g\lambda}{2\pi}\right) \text{th}\left(\frac{2\pi H}{\lambda}\right)$ we can find its wavelength in an independent way (Fig. 1c). Within the limits of Russell's estimations we get *H* = λ.

Thus, we conclude that Russell has come up against a critical phenomenon in which the energy of water motion associated with the boat is transformed into that of a free soliton. To the bifurcation corresponds the boat's speed $V_R = \sqrt{\frac{gH}{2\pi}}$ at which the boat becomes to advance its own ship wave. Since this is impossible the wave must reconstruct from $\vartheta = 0$ to $\vartheta > 0$. The break-point velocity accords with a boundary of the wave propagation named by "deep water" (*11*) (Fig. 2). More detailed mutual adjustment with Russell's data only slightly changes the resulted values (*12*).

Now it seems to be possible to inquire into some mysterious phenomena occurring in the World Ocean. According (*13*), there are numerous records about collisions of the ships with underwater rocks in the areas where, certainly, there can be no obstacles for navigation. These incidents have been attributed to an action of powerful elastic waves produced by underwater earthquakes, but this hypothesis has a serious disadvantage. In a great number of these events the shock waves should come from every possible direction, including from the stern. In such a situation, a ship should be accelerated rather then decelerated. We offer an alternative explanation. On the Ocean there are banks where the Russell's phenomenon can be displayed. A sharp ship deceleration may be easily taken for collision with an obstacle. For example, a 20 knot-going liner can run a risk of a 'soliton attack' from the depth of 68 m.

Recently, there has been much interest in investigating the nonlinear formation of freak or killer waves responsible for the loss of many ships and human lives in the Ocean (*14-16*). The freak wave is a localized water excitation, appearing from "nowhere" and reaching full height in a very short time, less than ten periods of surrounding waves. In the most advanced models a freak wave appears as a result of development of modulation instability (*17-19*). Not denying such opportunity we assume that in some cases a huge wave can be generated by means of the mechanism considered above. Participation of the ship is not required, since wave movement in the open sea is created by a wind. A spectrum of ocean waves strongly varies depending on the changing the wind force and direction. As a rule, the long waves leaving areas of storms or counterflows have the greatest energy. At pass above a site concerning an equal bottom the wavelength of a wave chain can appear close to depth that corresponds to a condition of "Russell-type instability" development. Then one crest from this chain will build up during the several periods at the expense of a neighbor's reduction. For typical waves of a ripple with the period of 15 seconds the critical depth is about third of kilometer. It is obvious, that the same effect can be reached in case of small bottom slopes at a respective alteration of wave parameters.

In the reservoirs of uniform depth one can observe a related ship-induced phenomenon. Russell knew about it, but admiral Alexey N. Krylov, the founder of the modern ship theory, has given a more pithy description (*20*).

The wave described by Krylov is also soliton. It aroused as a result of separation of a satellite wave from the ship. As the ship's speed approaches a critical value $\sqrt{gH}$ the satellite wave increases and slows the movement. Many popular books describe this phenomenon while an underlying mechanism is not studied. According to Krylov the satellite wave can exist in a speed range of $(0.75 \div 1.25)\sqrt{gH}$. Our analysis shows that it arises much earlier, after the transition by the ship the 'Russell's velocity' $\sqrt{\dfrac{gH}{2\pi}}$ (see Fig. 2). It suggests a general nature of both phenomena. Leaving a bank by the ship or, in the case of a canal, a sharp changing of its width promotes the separation of the wave. Then it may transform into another type of soliton, which we shall name the Krylov's one (Fig. 3).

In a reservoir of uniform depth RS can be induced within a very narrow interval of ship's speeds in the vicinity of the value $V_R$. A time is required that the process has been completed therefore, a limit value of the acceleration (deceleration) $a_R$ should not be exceeded. It is such a motion mode that has led to the boat's stop observed by Russell. The limit acceleration can be estimated with the help of dimensional considerations as $a_R \sim g \frac{\sqrt{gH}}{c_s}$, where $c_s$ is the sound velocity in water. It is rather difficult to satisfy this requirements having no prior knowledge of them. This circumstance explains the reason for the failures in the RS recreation. In particular, similar attempt has been undertaken in the course of the Edinburgh Conference on nonlinear waves in 1995. The speed of their boat was approximately $\sqrt{gH}$ while the arising soliton had a form of the Krylov's one (compare Fig. 3 and a photo from (*21*)). The absence of a reverse impact is additional evidence that the Krylov's soliton was reproduced rather than RS.

Thus, the water motion induced by an accelerating ship in a reservoir of uniform depth $H$ has a bifurcation at the ship's speed of $V_R = \sqrt{\frac{gH}{2\pi}}$. Near this point the wave pattern should be rearranged that can lead to the critical phenomena of two types (Fig. 4). In case of a small enough acceleration the ship will undergo a non-elastic impact, as though having come across an underwater obstacle, and it gives rise to the formation of a dome-like soliton. It is this surprising form of the water surface that explains the pathos of the Russell's report. At a faster passage the ship clings a satellite wave. It increases as the ship's speed approaches a total velocity limit $\sqrt{gH}$ for surface gravity waves in a reservoir of a uniform depth. The change of the motion mode, for example, of the ship's acceleration, reservoir's depth or channel's width, may promote the decoupling of this wave with its following transformation into a shaft-like soliton. At further accelerations any of these processes has not enough time to develop. Note that the theory, as for instance, using the Cauchy's problem approach for the incompressible potential flow, does yet not describe these hydrodynamical instabilities.

Though sound waves, unlike waves in liquids, do not possess the dispersion the bifurcation point $V_R$ for ships is apparently analogous to the sound velocity for flying machines. Instructions enact to pilots, for safety, to overcome the sound barrier on a forcing. One of features associated with the supersonic transition is the production of a sudden visible water-vapor cloud around an airplane. This condensation effect is commonly treated in terms of the Prandtl-Glauert singularity (*22*) however, it adds little to understanding the phenomenon. Our analysis shows that is more perspective to consider it as a manifestation of the Russell-type instability: at a sufficiently slow transition a plane can receive an explosion-like aerodynamic impact.

The picture of the wave field accompanying a moving charge, the bremsstrahlung being excluded, is rearranging also as the transition goes on from "underlight" motion to "superlight" one (*23*), when the Cherenkov's radiation of electromagnetic waves takes place (*24*). This suggests the existence of a similar light barrier for charge particles. Near the corresponding bifurcation it is possible to expect the occurrence of abnormal effects; some intimation on such feasibility is in (*25*). The study of barrier effects as a total unified class can contribute to the decision of some topical problems of hydrodynamics and electrodynamics.

**References and Notes**

1. J.S. Russell, in *Report of the 14th Meeting of the British Association for the Advancement of Science* (Jon Murray, London, 1844), pp. 311-390. "I was observing the motion of a boat which rapidly drawn along a narrow channel by a pair of horses, when the boat suddenly stopped – not so the mass of water in the channel which it had put in the motion; it accumulated round the prow of the vessel in a state of violent agitation, then suddenly leaving it behind, rolled forward with great velocity assuming the form of a large solitary elevation, a rounded, smooth and well defined heap of water which continued its course apparently without change of form or diminution of speed. I followed it on horseback, and overtook it still rolling at a rate of some eight to nine miles per hour, preserving its original figure some thirty feet long and a foot to a foot and a half in height. Its height gradually diminished and after a chase of one or two miles I lost it in the windings of the channel. Such, in the month of August 1834, was my first chance interview with the singular and beautiful phenomenon, which I have called the Great Wave of Translation".

2. A. S. Davydov, *Solitons in Molecular Systems* (Kluwer, Dordrecht, 1985).

3. C.H. Gu, *Soliton Theory and Its Application* (Springer-Verlag, New York, 1995).

4. A.T. Filippov, *The Versatile Soliton* (Birkhäuser, Boston, 1996).

5. M. Remoissent, *Waves Called Solitons: Concepts and Experiments, 3nd ed.* (Springer, New York, 1999).

6. E. Infeld, G. Rowlands, *Nonlinear Waves, Solitons, and Chaos, 2nd ed.* (Cambridge University Press, Cambridge, 2000).

7. J.L. Hammack, *Journal of Fluid Mechanics* **60,** 769 (1973).

8. J.L. Hammack, H. Segur, *Journal of Fluid Mechanics* **65**, 289 (1974).

9. J.L. Bona, W.G. Pritchard, L.R. Scott, *Phil. Trans. Roy. Soc. London* **302**, 457 (1981).

10. D.J. Korteweg, G. de Vries, *Phil. Mag* **39**, 422 (1895).

11. J. Lighthill, *Waves in Fluids* (Cambridge University Press, Cambridge, 2001).

12. Refined values of Russell's observation parameters: the Union Canal's depth ≈ 1.7 m, the boat's speed ≈ 3.2 miles per hour and the acceleration ≤ 0.03 m s$^{-2}$, time of it's stopping (development of instability) ≈ 3 second (2-3 wave periods), the soliton velocity ≈ 8.7 miles per hour.

13. V.V. Shouleykin, *Physics of Sea, 4-th ed.* (Nauka, Moscow, 1968).

14. A. R. Osborne, M. Onorato, M. Serio, *Phys. Lett. A* **275**, 386 (2000).

15. M. Onorato, A. R. Osborne, M. Serio, D. Resio, A. Pushkarev, V. E. Zakharov, C. Brandini, *Phys. Rev. Lett*. **89,** 144501 (2002).

16. J. L. Hammack, D. M. Henderson, H. Segur, *J. Fluid Mech.* **532**, 1 (2005)

17. V.E. Zakharov, A.I.Dyachenko, A.O.Prokofiev *Eur. J. Mech., B Fluids* **25,** 677 (2006).

18. M. Onorato A. R. Osborne, M. Serio, *Phys. Rev. Lett*. **96**, 014503 (2006).

19. P. K. Shukla, I. Kourakis, B. Eliasson, M. Marklund, L. Stenflo, *Phys. Rev. Lett*. **97,** 094501 (2006).

20. A.N. Krylov, in *Bulletin of Scientific & Technological Committee* (Moscow, 1931). Here is how he has described a case with a ship of the Russian Navy in vicinity of the Finland coast. "In 1912 a torpedo boat "Novik" at a speed of 20 knots, was passing a lighthouse (the distance was about 6 miles) at the entrance to one of the Finland skerries fairways. Near the lighthouse there was a wooden wharf resting on piles, and the wharf dais was 9 feet above the water level. It was dead calm. At the wharf there was a boat (keel up), and two boys of 10 and 6 years of age were playing near this boat. The elder boy saw a large wave nearing the wharf, and ran in the direction from the water while the younger boy remained on-site. The wave reached the wharf and washed away the boat and everything that was on the wharf, and the boy was washed away as well and drowned. Nobody could see this from the "Novik". Only when they came to Gang the commander got a telegram about the accident. An investigation has been started, and the Navy Minister trusted me to make report on this matter. It so happened that on the way of the "Novik" there was a short bank of the depth of 35 feet. Such a depth turns to be a 'critical' one ($\sqrt{gH}$) for the speed of 20 knots. And just on this bank a huge wave was formed, which ran farther on, and caused the troubles. This was indeed an unforeseen event on sea" (*26*).

21. A resume of the conference "Coherent Structures in Physics and Biology", Edinburgh, Scotland, July 1995, with the photo of the participants trying to reproduce Scott Russell's solitary wave: *Nature* **376**, 373 (1995).

22. H. Glauert, *The Elements of Aerofoil and Airscrew Theory* (Cambridge University Press, Cambridge, 1948).

23. G. Afanasiev, V. Kartavenko, E. Magar, *Physica B* **269**, 95 (1999).

24. J. V. Jelley, *Cerenkov Radiation and Its Applications* (Pergamon Press, Oxford, 1958).

25. A.S. Vodopianov, V.P. Zrelov, A.A. Tyapkin, *Particles and Nuclei* **2**(99), 35 (2000).

26. A.N. Krylov, *My Reminiscences* (Sudostroenie, Leningrad 1984) (in Russian).

27. We are grateful to C. Eilbeck (Heriot-Watt University, Edinburgh) for help in investigation, E.A. Kulikov and A.B. Rabinovich from Oceanology Institute (Moscow) for useful discussions, and D. Hazelwood for habituation with experimental data.
**Fig. 1.** Determination of needed data lacking in the Russell's report. Zones of the most plausible values are shown by rectangles. (**A**) The canal depth $H$ versus soliton velocity $v$ for three values of soliton height $h$ in feet; the boldface line labels an interval reported by Russell. (**B**) Limit wavelength $\lambda_l$ of ship-generated wave perturbations versus ship speed $V$. (**C**) The ship wave phase velocity versus wavelength. The latter is found independently from the surface gravity wave dispersion law.

**Fig. 2.** The interrelation ship-to- ship wave. The ship's speed / the wave velocity squared versus the wavelength (in normalized units). The curve 1 describes the dispersion of surface gravity waves; the curve 2 relates the ship's speed and a long wavelength limit of the ship-generated wave perturbations. The position and length of the red line determine a zone of the Russell's critical phenomenon; the blue band marks an area of the Krylov's one. The color intension correlates with the height of a satellite wave.

**Fig. 3.** A schematic sketch of the ship-induced soliton form and pattern preceding the soliton formation: (**A**) for the Russell's phenomenon; (**B**) for a phenomenon of the Krylov's type.

**Fig. 4.** A dynamical phase diagram of the ship associated critical phenomena in coordinates 'speed- acceleration of a ship'. **R** is the Russell's critical phenomenon zone (red), **K** is the Krylov's one (green); a pre-critical zone (blue), a post-critical one (yellow).

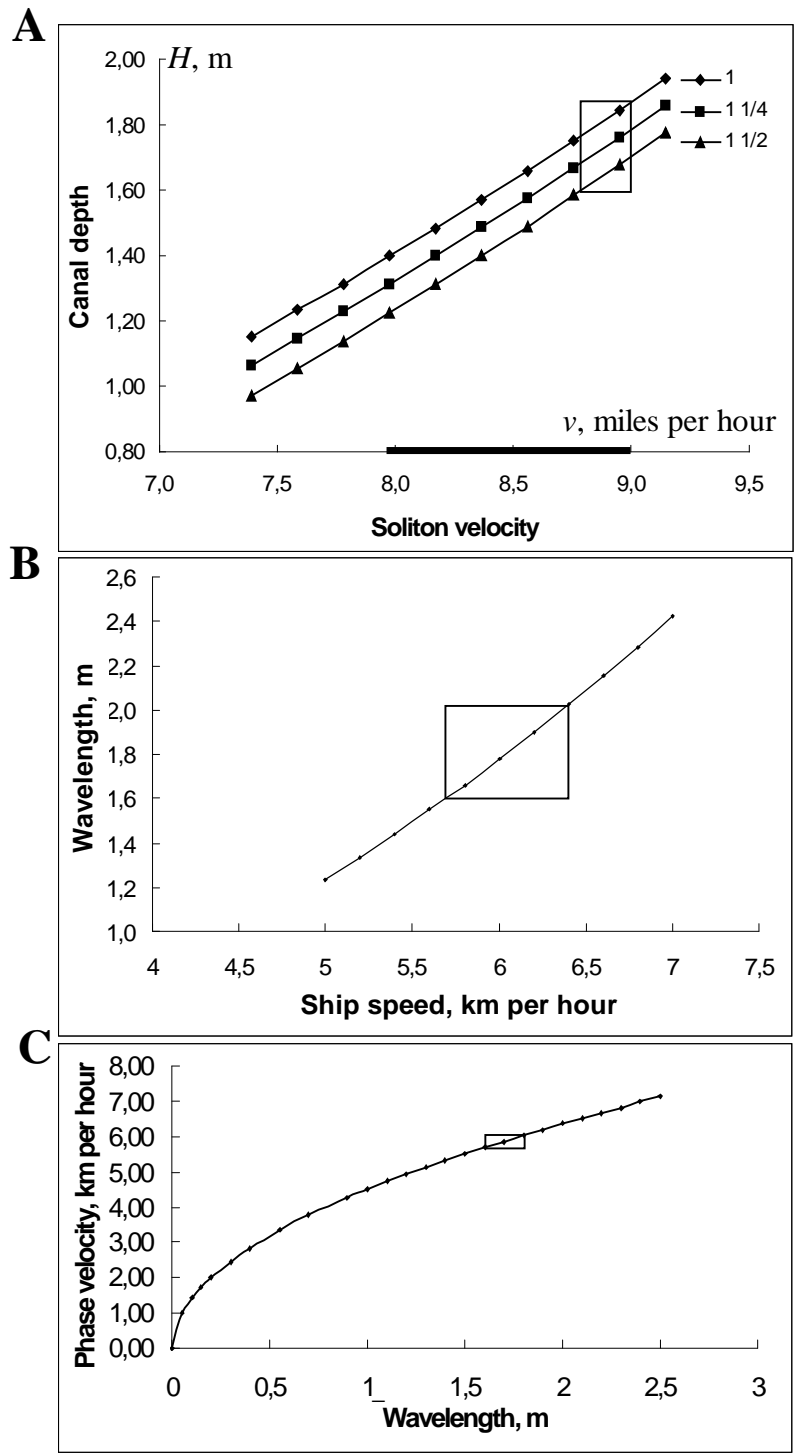

Figure 1.

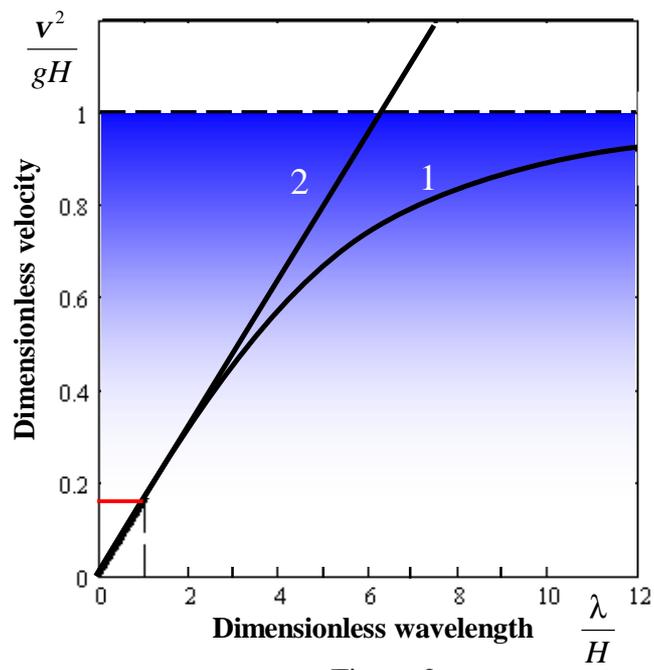

Figure 2.

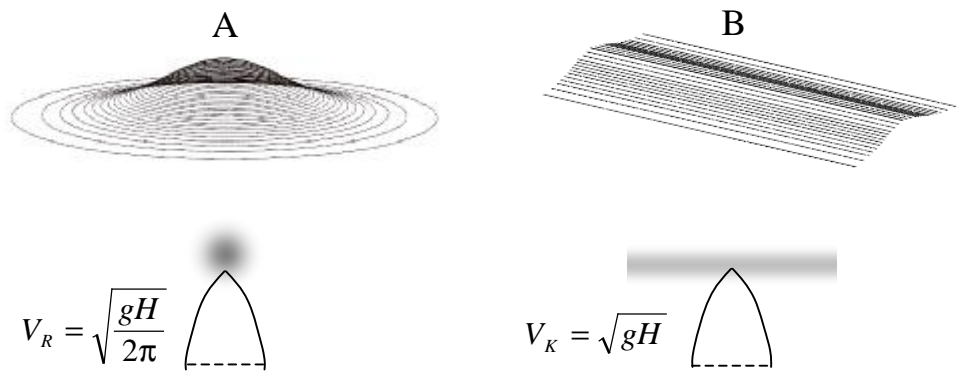

Figure 3.

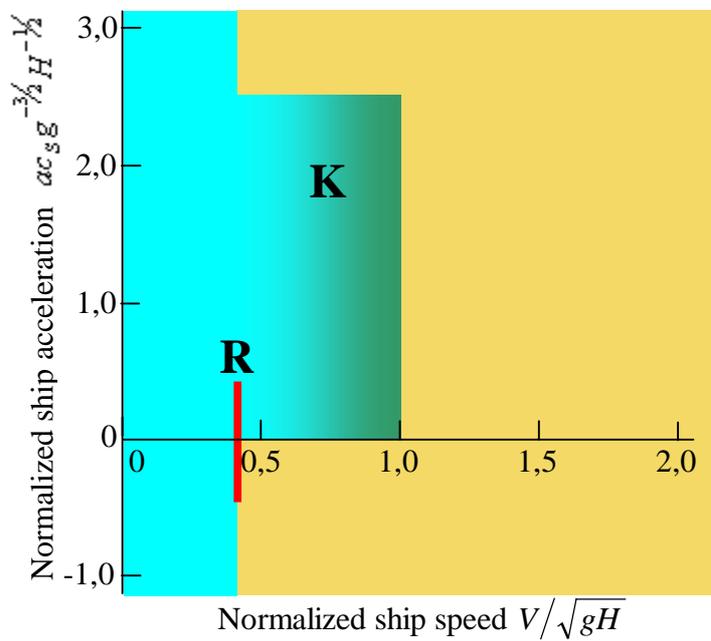

Figure 4.